\documentclass[compsoc, twoside]{IEEEtran}
\usepackage{ifpdf}
\ifCLASSINFOpdf
\usepackage[pdftex]{graphicx}
\usepackage{nicefrac}
\DeclareGraphicsExtensions{.pdf,.jpeg,.png}
\else
\DeclareGraphicsExtensions{.pdf}
\fi
\usepackage[cmex10]{amsmath}
\usepackage{array}
\usepackage{cite}
\usepackage{amssymb}  
\usepackage{balance} 
\usepackage[papersize={19cm,26.5cm},top=1cm,left=1cm,right=1cm,bottom=1cm,headheight=18pt,headsep=2mm,footskip=4mm]{geometry}
\usepackage[footnotesize]{caption}
\usepackage{tabularx}
\usepackage{booktabs}
\usepackage{siunitx}
\usepackage{subcaption}
\usepackage{ctable}
\usepackage[T1]{fontenc}
\usepackage[utf8]{inputenc}
\usepackage{enumitem}
\usepackage{multirow}
\usepackage{ragged2e}
\definecolor{Gray}{gray}{0.9}
\usepackage{makecell, cellspace, caption}
\DeclareFontFamily{U}{matha}{\hyphenchar\font45}
\DeclareFontShape{U}{matha}{m}{n}{
      <5> <6> <7> <8> <9> <10> gen * matha
      <10.95> matha10 <12> <14.4> <17.28> <20.74> <24.88> matha12
      }{}
\DeclareSymbolFont{matha}{U}{matha}{m}{n}

\DeclareMathSymbol{\Lt}{3}{matha}{"CE}
\DeclareMathSymbol{\Gt}{3}{matha}{"CF}

\graphicspath{ {./images/} }

\DeclareUnicodeCharacter{FB01}{fi}
\makeatletter
\newcommand\subparagraph{%
	\@startsection{subparagraph}{5}
	{\parindent}
	{3.25ex \@plus 1ex \@minus .2ex}
	{-1em}
	{\normalfont\normalsize\bfseries}}
\makeatother

\usepackage{titlesec}
\usepackage{indentfirst}
\titleformat{\section}{\sffamily\bfseries}{\thesection.}{.4em}{}
\titleformat{\subsection}{\sffamily\bfseries\small}{\thesubsection.}{.4em}{}
\titleformat{\subsubsection}{\sffamily\itshape\small}{\thesubsubsection.}{.4em}{}

\makeatletter
\def\ScaleIfNeeded{%
	\ifdim\Gin@nat@width>\linewidth
	\linewidth
	\else
	\Gin@nat@width
	\fi
}
\makeatother

\allowdisplaybreaks

\begin{document}
\title{An Analysis of Uplink Success Probability in Multi-Cell Lora Networks Under Different Channel Models}

\author{Tien Hoa Nguyen*\thanks{\hspace{5mm}Tien Hoa Nguyen is with the school of electronic and electrical engineering (SEEE), Hanoi University of Science and Technology, Vietnam.\vspace{3pt}},
Van Dai Do\thanks{\hspace{5mm}Van Dai Do is with the school of electronic and electrical engineering (SEEE), Hanoi University of Science and Technology, Vietnam. \vspace{3pt}}
\thanks{\protect\\ *Corresponding author: Tien Hoa Nguyen (e-mail: hoa.nguyentien@hust.edu.vn)
	
	\noindent Manuscript received December 6, 2021; revised December 31, xxxx; accepted March 07, xxxx.
	
	\noindent Digital Object Identifier 10.31130/jst-ud.xxxx.xxx}}

\makeatletter
\newcommand\ShortTitle{Clustering Internet of Things: A Review}
\newcommand\Authors{Sahil Sholla \textit{et al.}}
\makeatother

\markboth{\scalebox{0.98}{UD - JOURNAL OF SCIENCE AND TECHNOLOGY: ISSUE ON INFORMATION AND COMMUNICATIONS TECHNOLOGY, VOL. xx, NO. xx, xxxx}}{\Authors : \ShortTitle}

\IEEEpubid{ISSN 1859-1531}

\IEEEtitleabstractindextext{%
\begin{abstract}
\justifying	
The development of the low power wide area network (LPWAN) for the internet of things (IoTs) is expected to grow widely, allowing remote monitoring of smart devices from a distance of up to several kilometers. This paper studies the performance and success probability of multi-cell LoRa networks. Using tools of stochastic geometry, the paper analyzes the important metric namely success probability in both Rayleigh and Rician channel models. The obtained analysis helps investigate and evaluate other quality criteria in the multi-cell LoRa network such as throughput, SNR and SIR requirements. Moreover, we provide numerical simulation results to corroborate the theoretical analysis and to verify how our analysis can characterize the given reliability target.
\end{abstract}

\begin{IEEEkeywords}
LoRa, LoRaWAN, Probability, Rayleigh, Rice, SNR, SIR.\\
\end{IEEEkeywords}}

\maketitle

\setcounter{page}{1}

\IEEEraisesectionheading{\section{Introduction}}
\IEEEPARstart{T}{he} positive impacts of IoT on society, the environment and the industry are expected to be significant, with billions of devices that are connected relating to IoT \cite{Qadir2021}. There are plenty of reasons why IoT plays a key role in our lives; however, there are two main ones, which are the connectivity and affordability of wireless infrastructures \cite{Yu2020, Hoeller2018} LPWAN technologies have drawn much interest in this industry for its main features: interconnecting battery-powered devices with low-bandwidth, low bit rates over long ranges. Together with other LPWAN technologies (SigFox, Weightless-W, etc.), it proposes connectivity up to tens of kilometers for low data rate, low power and low throughput applications. The market for this technology is expected to be huge, mainly in smart cities and smart agricultural projects \cite{Vogelgesang2021, Georgiou2020}.

LoRa networks and their applications have been deployed in many countries; however, there is one aspect of LoRa that has not been studied adequately, which is the oversimplification of non-interactable and independent gateways \cite{Beltramelli2018}. In addition, there have also not been enough comparisons to evaluate the effectiveness for different fading models. Those might lead to incorrect conclusions in papers and false optimize protocols \cite{Mahmood2019, Nguyen2020}.

In a context in smart cities with a large number of connected measuring devices, the exploitation of the LoRa system depends on the number of devices supported and its scalability \cite{Tu2020}. In addition, devices are often clustered at gateways, where the interference models such as co-cell and intra-cell of a multi-cell network must be defined \cite{Afisiadis2020}. This stems from the imperfect orthogonality between the different \emph{SFs}. In addition, transmission from different \emph{SFs} does not eliminate interference with neighboring \emph{SFs}, so LoRa systems need to require a certain level of Signal-to-interference (SIR) protection \cite{Kim2020}. In the presence of SIR, the performance of the LoRa system depends on specific signal to noise ratios (SNR) at the \emph{SFs}. In this respect, SNR and SIR are important metrics of evaluating the LoRa system \cite{Reynders2018}.

In this paper, our main contribution is to focus on analyzing the success probability in two different fading models, namely Rayleigh and Rician, hence comparing their performances under the same configuration. Earlier works widely used Rayleigh fading model \cite{Georgiou2017} while not considering Rician fading model, which are clearly stimulated and illustrated in our paper. Both models stimulated in our paper also do allow gateways to interact within their ranges, and clearly illustrate the traceability of them.

In addition, we argue that elements in LoRa networks, which are the gateways and the end devices, can be approximately described by inhomogeneous Poisson point processes (PPPs). Then, we obtain closed form expressions of many network features. Finally, we compare the probability between the two models to find whether there are any significant differences between them under the same conditions.

Notations: Vectors and Matrices are denoted by boldface small and big letters, correspondingly. The superscripts T and H stand for the transpose and conjugate transpose. $I_K$ is the $K \times K$ identity matrix. The operator $E{.}$ is the expectation of a random variable. The notation $||.||$ for the Euclidean norm. 

\section{LoRa standardization and Motivation}

In this part, we briefly discuss the related standardization of LoRa and LoRaWAN. LoRa is known as a spread spectrum modulation technique, working based on chirp spectrum technology. LoRa operates in sub-gigahertz radio frequency bands. For each region, LoRa has different working ranges. In Europe, it works according to EU868 (863-870/873 MHz), while in Asia, it follows AS923 (915-928 MHz). LoRa systems include gateways (GWs), end-devices (EDs) and the NetServer. These elements form a star of stars topology with NetServer at the root, GWs at level one and EDs as the leaves. Tab.~\ref{tab:1} illustrates the main features of LoRa, specifically for a 25-byte message.

\begin{table*}[htbp]
	\caption{LoRa Characteristics of a 25-Byte message at BW = 125 kHz}
	\centering
	\begin{tabular}{|c|c|c|c|c|c|c|}
		\toprule
		\textbf{SF} & \textbf{bit-rate (kb/s)} & 		\textbf{Packet air time  (ms)} & 
		\textbf{Tx/h} &
		\textbf{Receiver Sens (dBm)} &
		\textbf{SNR} $q_{SF}$(dBm) & 
		\textbf{Range} $q_{SF}$(km) \\
		\midrule
		7 & 5.47 & 36.6 & 98 & -123 & -6 & $d_0$ - $d_1$ \\
		\midrule
		8 & 3.13 & 64 & 56 & -126 & -9 & $d_1$ - $d_2$ \\
		\midrule
		9 & 1.76 & 113 & 31 & -129 & -12 & $d_2$ - $d_3$ \\
		\midrule
		10 & 0.98 & 204 & 17 & -132 & -15 & $d_3$ - $d_4$ \\
		\midrule
		11 & 0.54 & 372 & 9 & -134.5 & -17.5 & $d_4$ - $d_5$ \\
		\midrule
		12 & 0.29 & 682 & 5 & -137 & -20 & $d_5$ - $d_6$ \\
		\bottomrule
	\end{tabular}%
	\label{tab:1}%
\end{table*}%

The heart of the technology, chirp spread spectrum (CSS) modulation, ensures adaptive data rates and lets the system trade-off throughput for different properties, such as coverage range, or energy consumption, while maintaining the same bandwidth (BW). This process is often done by the Net Server, which regulates the BW and the SF. LoRa modulation has a total of 6 SF from SF7 to SF12, which controls the chirp symbol $T_s = \nicefrac{2^{SF}}{\rm BW}$. For that reason, the time-on-air of a transmission using LoRa increases exponentially with SF (see Tab.~\ref{tab:1}).

\begin{table}[htbp]
	\caption{SIR collision thresholds in dB between different SFs}
	\centering
	\begin{tabular}{|c|c|c|c|c|c|c|}
		\toprule
		\textbf{SF} & 
		\textbf{7} & 
		\textbf{8} & 
		\textbf{9} & 
		\textbf{10} & 
		\textbf{11} & 
		\textbf{12} \\
		\midrule
		7 & 1 & -8 & -9 & -9 & -9 & -9 \\
		\midrule
		8 & -11 & 1 & -11 & -12 & -13 & -13 \\
		\midrule
		9 & -15 & -13 & 1 & -13 & -14 & -15 \\
		\midrule
		10 & -19 & -18 & -17 & 1 & -17 & -18 \\
		\midrule
		11 & -22 & -22 & -21 & -20 & 1 & -20 \\
		\midrule
		12 & -25 & -25 & -25 & -24 & -23 & 1 \\
		\bottomrule
	\end{tabular}%
	\label{tab:2}%
\end{table}%

LoRaWAN is an open protocol letting devices use LoRa for communication. It makes use of the pseudo-orthogonality between SFs to serve more EDs. Tab.~\ref{tab:2} shows the SIR threshold needed to successfully send a packet between different SFs. While LoRa is in the physical layer, LoRaWAN is in the datalink and network layers. As mentioned in the introduction, LoRaWAN key attributes are long-range, low-power, and low data rates. The range of communication mostly depends on whether the Line-of-Sight (LoS) channel is available or not. In places where there are any objects that create non-Line-of-Sight (nLoS), such as the building blocks in the cities, the communication distance in fact can be much shorter than 10 km. LoRaWAN protocol follows the Aloha method, which lets devices communicate only when there is data ready to be sent. Since there is no need to synchronize, a device can go back to silent mode after sending a packet. This contributes to the low-power characteristic of LoRaWAN. The data rates also vary from region to region. In Europe, for example, the data rate range is from 250bps to 5.5kbps. This is considered low for daily activities, for instance, surfing the internet or watching movies. However, it meets the requirements for tiny amounts of data, from simple devices such as sensors. This is where LoRa and LoRaWAN really stand out from other means of wireless communications.

\section{SIR and SNR analysis for uplink model in multi-gateway}

\subsection{Spatial Distribution of GWs and EDs}

We suppose GWs and EDs are described by the homogeneous Poisson point processes GW and ED, with constant intensity functions $\phi_{\rm GW}$ and $\phi_{\rm ED}$, respectively [5]. Since the number of ED is much greater than the number of GW in our simulations, we have $\delta_{\rm GW} \ll \delta_{\rm ED}$. Each ED and GW is described as a point $x_i$ and $y_i$ in the PPPs, respectively. In addition, our coordinate system is supposed to have ED $i=0$ at the origin to simplify calculations. $d_{ij}= |x_i - y_j|$ is the Euclidean distance in km from $i^{th}$ ED to $j^{th}$ GW.

\subsection{Rayleigh and Rician Fading Models}

In the state-of-the-art studies, where the authors in \cite{Georgiou2017} just consider multi-gateway interactions of LoRa networks under Rayleigh channel fading.  However, it is obvious that Rayleigh fading is suitable for rich scatter environments, it, as a result, is the best suit for the urban and/or indoor environments where LoS is generally hard to exist. The Rician fading model, on the other hand, is employed when a strong LoS path exists, thus, is typically used in the outskirts and/or rural areas. Additionally, it is noted that LoRa can be applied to either the urban or rural areas. As a consequence, we investigate the performance of LoRa networks under both fading distributions.

In the present paper, two different fading distributions are taken into account, namely, the Rayleigh and Rician distributions. It is noted that Rayleigh distribution is employed in the urban and/or indoor environments where LoS hardly exists. Rician distribution, on the other hand, is typically used in rural areas where the received signal is dominated by a strong LoS path. In the former circumstance, we model the channel gain $h_{ij}^2$ between $i^{th}$ ED and $j^{th}$ GW by an exponential random variable of mean 1, assuming that our channel $h$ (quasi-static) here is modeled as an independent, circularly symmetric, zero-mean complex Gaussian random variable with unit variance. In the latter circumstance, the channel gain $h_{ij}^2$ is described as a Rician variable of mean 1, with its variance is also 1. We do not include the effects of log-normal shadowing in this paper, since such fluctuations are not expected to significantly change the results of our qualitative analysis. The white Gaussian noise (AWGN) in this model is assumed to have zero-mean and variance \cite{Tesfay2020}
\begin{equation}
    N({\rm dBm}) = -174 + 10 \log({\rm BW}) + {\rm NF},
\end{equation}
where the first term is the thermal noise in 1 Hz of bandwidth and can only be affected by changing the temperature of the receiver. NF is the fixed receiver noise figure, here it is taken to be 6 dBm due to the model's hardware implementation. For simplicity, we assume that uplink uses a BW = 125kHz channel and a 25 Byte packet. In addition, all end-devices are assumed to transmit with a constant power $\varepsilon = 19$dBm.

\subsection{Path Loss Attenuation}

Path loss is a major component in the analysis and design of a telecommunication system. In order to adjust the coverage and capacity of wireless networks, we pay attention to the behavior of the attenuation function. For simplicity, in this paper, we assume that the transmitted signal undergoes a path loss attenuation, described by the function $p(d_{ij})=\left(\frac{\lambda}{4 \pi d_{ij}} \right)^2$, derives from the Friis transmission equation, where $\lambda = 34.5 cm$ is the carrier wavelength, and $\eta \geq 2$ is the path loss exponent.

\subsection{Spreading Factor in different regions of Uplink Transmissions}

$SF$ in a LoRa network controls the speed of data transmission. Lower SF reduces the range of LoRa transmission, since the bit rate is increased, and processing gain is reduced. In this paper, we assume that each ED will transmit the $SF$ set by the distance between that ED and its nearest GW, according to Tab.1. For instance, if $d_{00}$ is in $d_2,d_3$, then ED $i^{th}=0$, it will transmit with $SF=9$. This allows our network to effectively divide EDs eight regions orbiting around each GW. By doing so, we can easily assign $SF$ to each ED and make sure that every ED has only one $SF$. It is important to note that $\phi_{\rm GW} << \phi_{\rm ED}$, to guarantee that there is also ED with higher $SF$ in the analysis.

\subsection{SNR and SIR requirements}

Usually, in wireless telecommunications models, signal-to-interference-plus-noise ratio (SINR) is used to represent path loss with distance and other factors, such as background noise, interfering strength of other transmission. Unlike other means of transmissions that use SINR to measure the effectiveness, LoRa has these two separate conditions of SNR and SIR, which we will further discuss below.

\subsubsection{SNR requirement}

The first condition is concerned with whether the received SNR is below the SF specific threshold $q_{SF}$ (see Tab.~\ref{tab:2}). The instantaneous SNR between ED ith and GW jth can be defined as
\begin{equation}
    {\rm SNR}_{ij} = \frac{\varepsilon |h_{ij}|^2}{Np(d_{ij})},
\end{equation}
where $\varepsilon$ and $N$ are the transmit power and noise variance which are defined in Section 3.2; $p(d_{ij})$ is the path-loss and defined in Section 3.3, and $h_{ij}^2$ is the channel gain between $i^{th}$ ED and $j^{th}$ GW modeled as described in Section 3.2. In the case of Rayleigh channel, the channel $h_{ij}^2$ is modeled by an exponential random variable of mean 1, whereas in the case of the Rician channel, it is described as a Rician variable of mean 1 and unit variance. The first condition can now be formulated as the complement of the connection probability
\begin{equation}
    H_1 = P[{\rm SNR}_{ij} \geq q_{SF}|d_{ij}],
\end{equation}
which captures the probability that at any instance of time, a received signal $s_1(t)$ from $i^{th}$ ED locating $d_{ij}$ km away from $j^{th}$ GW will not satisfy the SNR threshold $q_{SF}$, a piece-wise function of the distance $d_{ij}$.

\subsubsection{SIR requirement}

The second condition examines whether the signal-to-interference (SIR) between an ED and a GW is greater or equal than the SIR ratio threshold. As seen in Tab. 2, there is a same SIR ratio threshold for every co-SF collision $\tau = 1dB = 1.259$, while for non-co-SF collisions, that number varies from -8 to -25 dB (which corresponds to 0.15 to 0.0032). We can now define the SIR between $i^{th}$ ED and $j^{th}$ GW as
\begin{equation}
    {\rm SIR}_{ij} = \frac{\varepsilon|h_{ij}|^2}{p(d_{ij})} \frac{1}{I_j},
\end{equation}
where $I_j = \Sigma \frac{\psi_{ik} \varepsilon|h_{ij}|^2}{p(d_{ij})}$ where $k \neq i$ is the total interference of co-SF collisions at $j^{th}$. $\psi_{ik}$ is set to 1 if $k^{th}$ ED is having the same SF while transmitting with $i^{th}$ ED, and zero otherwise. 

It is noted that the SNR and SIR in the LoRa networks are identical for both uplink and downlink transmissions and are different from the cellular network, where power control is compulsory in all uplink transmissions. The second condition can now be formulated as
\begin{equation}
    H_2 = P[{\rm SIR}_{ij} \geq \tau | d_{ij}].
\end{equation}

\subsubsection{Probability of successful uplink transmission}

With the given conditions, we formulate the probability of a successful uplink transmission by ED $i^{th}$ as
\begin{equation}
    H(x_i) = P \left[ \cup \left[ \left( {\rm SNR}_{ij} \geq q_{SF_i} \right) \cap \left({\rm SNR}_{ij} \geq \tau \right) \right] \right],
\end{equation}
where the events $H_{1ij}= \left({\rm SNR}_{ij} \geq q_{SF_i} \right)$ and $H_{2ij}= \left({\rm SIR}_{ij} \geq \tau \right)$ are assumed to be independent to make the problem more mathematically tractable and more suitable for simulations. This probability is also used to evaluate the performance of Rayleigh channel and Rician channel. 

\section{Results and Discussions}

In this section the numerical simulation results to validate the theoretical analysis and verify our analysis will be presented. The simulation configurations are given in the Tab.~\ref{tab:3} as follows:

\begin{table}[htbp]
	\caption{The simulation parameters}
	\centering
	\begin{tabular}{|l|c|}
		\toprule
		Stimulation Radius & 
		20 km \\
		\midrule
		Gateway Intensity & 0.005 GW/${\rm km}^2$ \\
		\midrule
		End-device Intensity & 5 ED/${\rm km}^2$\\
		\midrule
		Bandwidth & 125 kHz \\
		\midrule
		ED Transmit Power $(\varepsilon)$ & 19 dBm \\
		\midrule
		Wavelength & $34.5 \times{10}^{-5}$ km \\
		\midrule
		Packet Size & 25 Byte \\
		\bottomrule
	\end{tabular}%
	\label{tab:3}%
\end{table}%

\subsection{SNR results}

SNR is the ratio of the useful signal power to the unwanted signal power, for example noise. To improve LoRa performance, this system often uses forward error correction (FEC) and spread factor ($SF$) that allows for significant SNR improvement. The $SF$ factor had the most significant impact on the LoRa system. Lower $SF$ reduces power consumption time, increases data rate.

\begin{figure}[!h] 
\centering
\begin{subfigure}{0.5\textwidth}
  \includegraphics[width=1\linewidth]{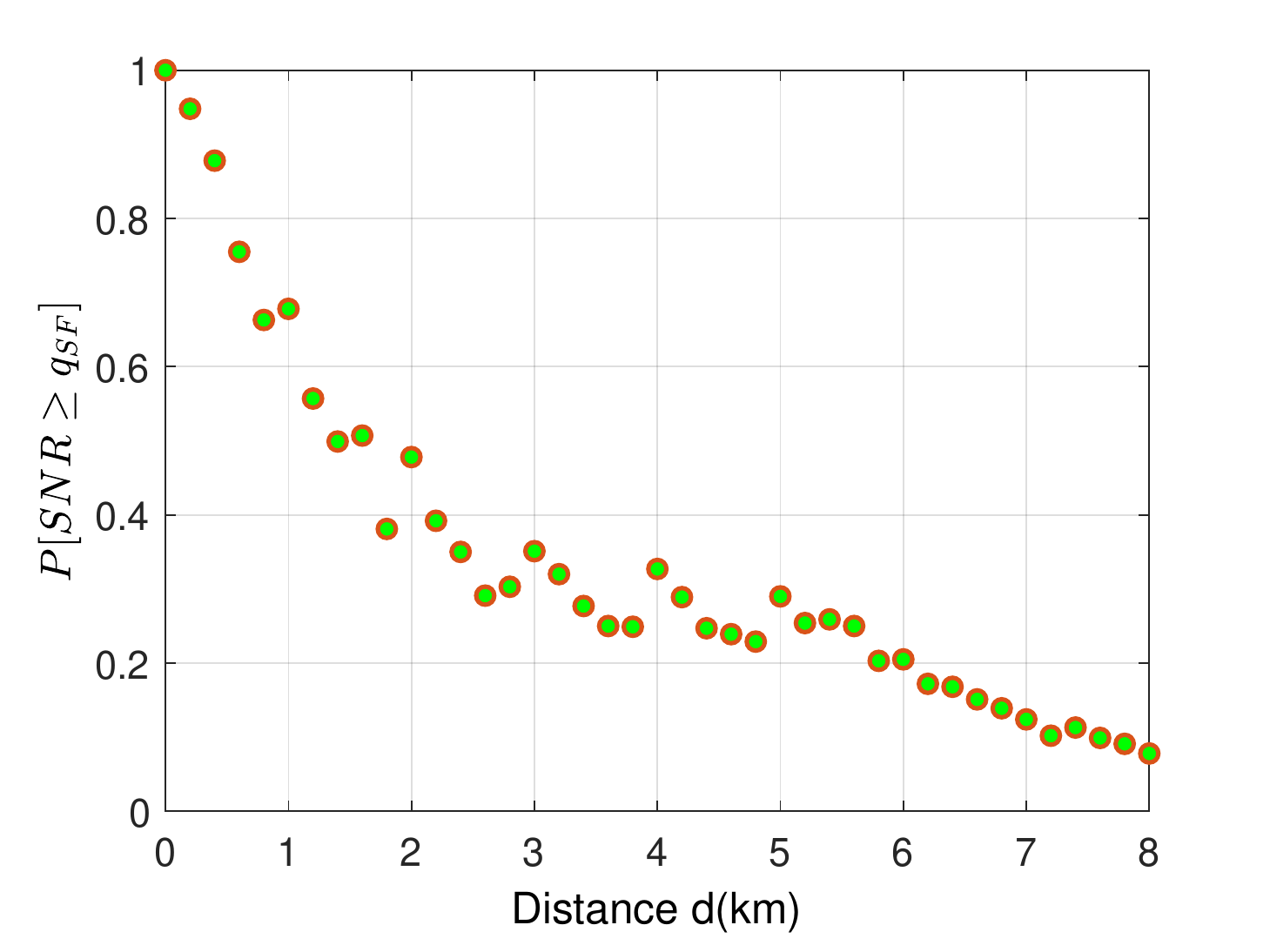}
  \caption{}
\end{subfigure}
\begin{subfigure}{0.5\textwidth}
  \includegraphics[width=1\linewidth]{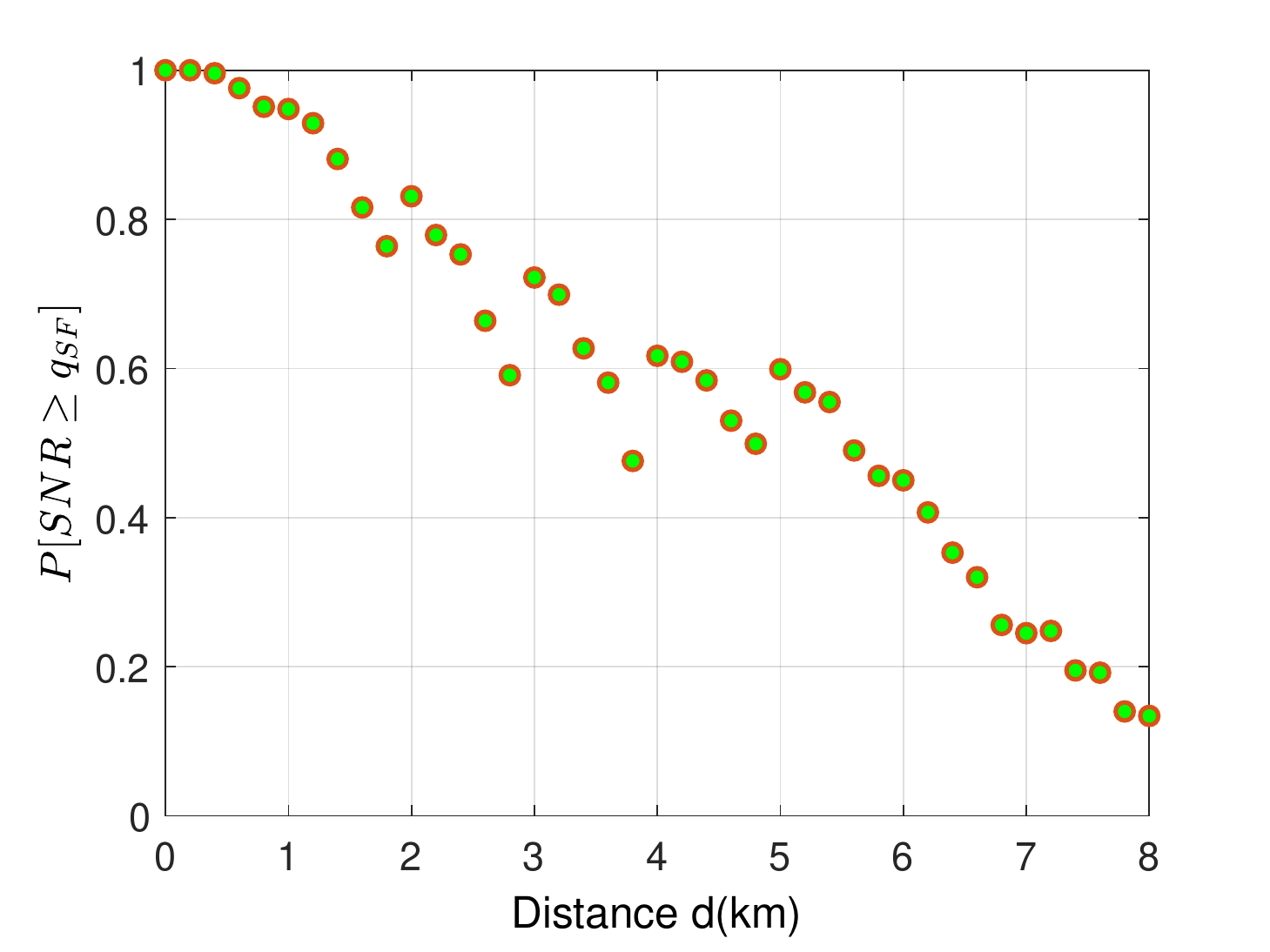}
  \caption{}
\end{subfigure}
\caption{Linear plot of simulation results of the probability $P[{\rm SNR}_{00}\geq q_0|d_{00}]$ between ${\rm ED} i^{th} = 0$ and its nearest ${\rm GW} j^{th} = 0$ for (a) Rayleigh channel and (b) Rician channel.}
\end{figure}

Fig.1a and Fig.1b illustrate Monte Carlo simulation results of the SNR condition of the two different channels. Overall, it is obvious that the probability of both cases decreases over time; however, the Rayleigh channel's probability decreases faster compared to Rician's. Over the range considered (8km), the probability of $H_1$ satisfying the Rician channel is higher than Reyleigh's. In contrast, EDs with lower $SF$ undergo different patterns for the two different channels. For the Rician channel, EDs having their closest GW within 1km obtain a success rate of more than 95\% as observed in Fig.1b. However, that probability of Rayleigh channel drops to just more than 65\%.

\subsection{SIR results}

The signal-to-interference ratio (SIR) is used to evaluate the performance of our models here. The ratio is the quotient between the mean received modulated carrier power and the mean received co-channel interference power from other transmitters. Inter-SF interference resulting from transmissions from neighboring $SFs$ is not orthogonal, so in LoRa systems a certain degree of SIR protection is required in addition to evaluating the SNR performance.

Fig.2 illustrates the simulation results of the SIR condition for the two channels. We observe a striking saw-tooth of the SIR condition, and the successful probability with respect to the transmission distance. In fact, this is a unique feature of LoRa and is the direct sequence of $q_{SF}$. The main reason is that when the transmission distance belongs to different $SF$ regions, we have a novel threshold that generally decreases with the increase of the transmission distance, thus, the probability first boosts up and then goes down for each region.
\begin{figure}[!h] 
\centering
\begin{subfigure}{0.5\textwidth}
  \includegraphics[width=1\linewidth]{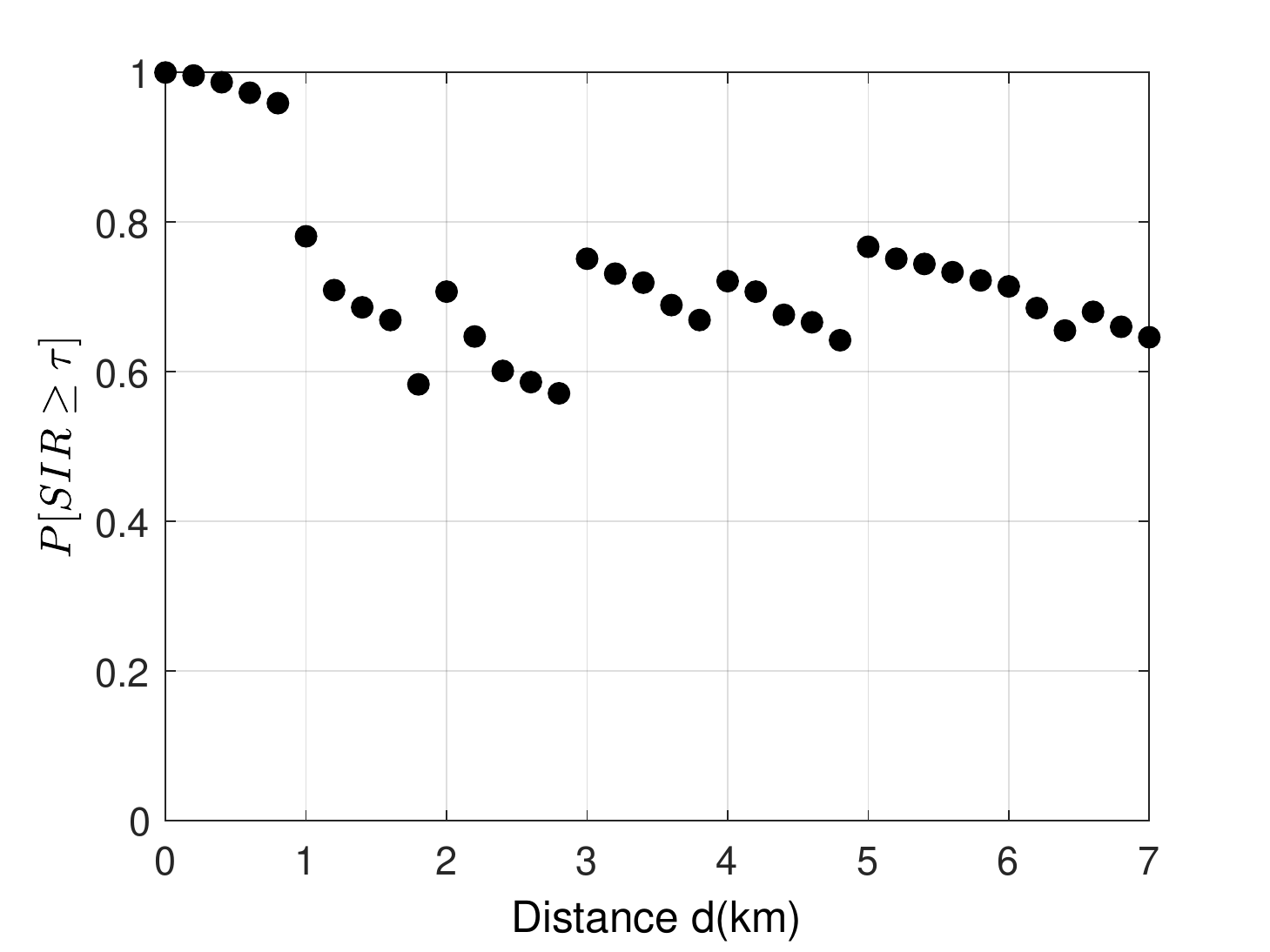}
  \caption{}
\end{subfigure}
\begin{subfigure}{0.5\textwidth}
  \includegraphics[width=1\linewidth]{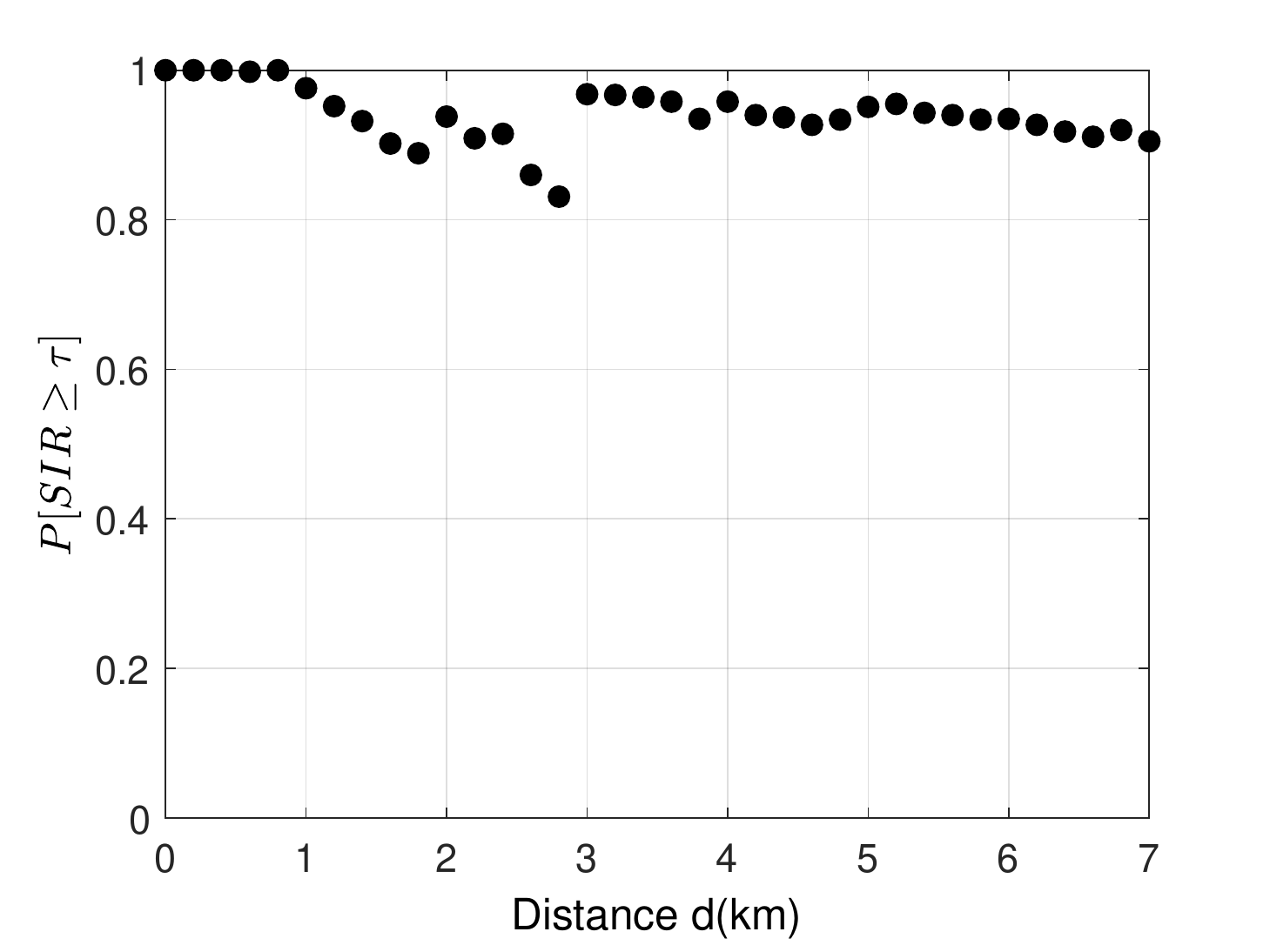}
  \caption{}
\end{subfigure}
\caption{Numerical simulations of $P[{\rm SIR}_{00}\geq\tau]$ as a function of $d_{00}$ for $\lambda_{\rm ED} =5{\rm ED}$ per km and $\lambda_{\rm GW} = 0.05{\rm GW}$ per km for (a) Rayleigh channel and (b) Rician channel}
\end{figure}

Looking at Fig.2a, in the first kilometer, Rayleigh channel also witnesses a high probability of meeting the SIR requirement; however, for the rest of the distance range, it stays in between just under 60 per cent and 80\%, which is considerably lower than Rician's. On the other hand, in Fig.2b, for the Rician channel, the rate of ED satisfying SIR requirement is high over the range considered. The rate maintains at nearly 100\% in the first kilometer, then it stays more than 80\% in the rest of the range. 

\subsection{Success transmission rate}

We now consider the success transmission rate, which is made by combining the SIR requirement and SNR requirement for two channels. As mentioned earlier in the explanation of Fig.2, the saw-tooth of the success transmission rate is a unique feature of LoRa. Overall, from Fig.3b, it is seen that the success rate of the Rician channel is higher than Rayleigh's. 

\begin{figure}[!h] 
\centering
\begin{subfigure}{0.5\textwidth}
  \includegraphics[width=1\linewidth]{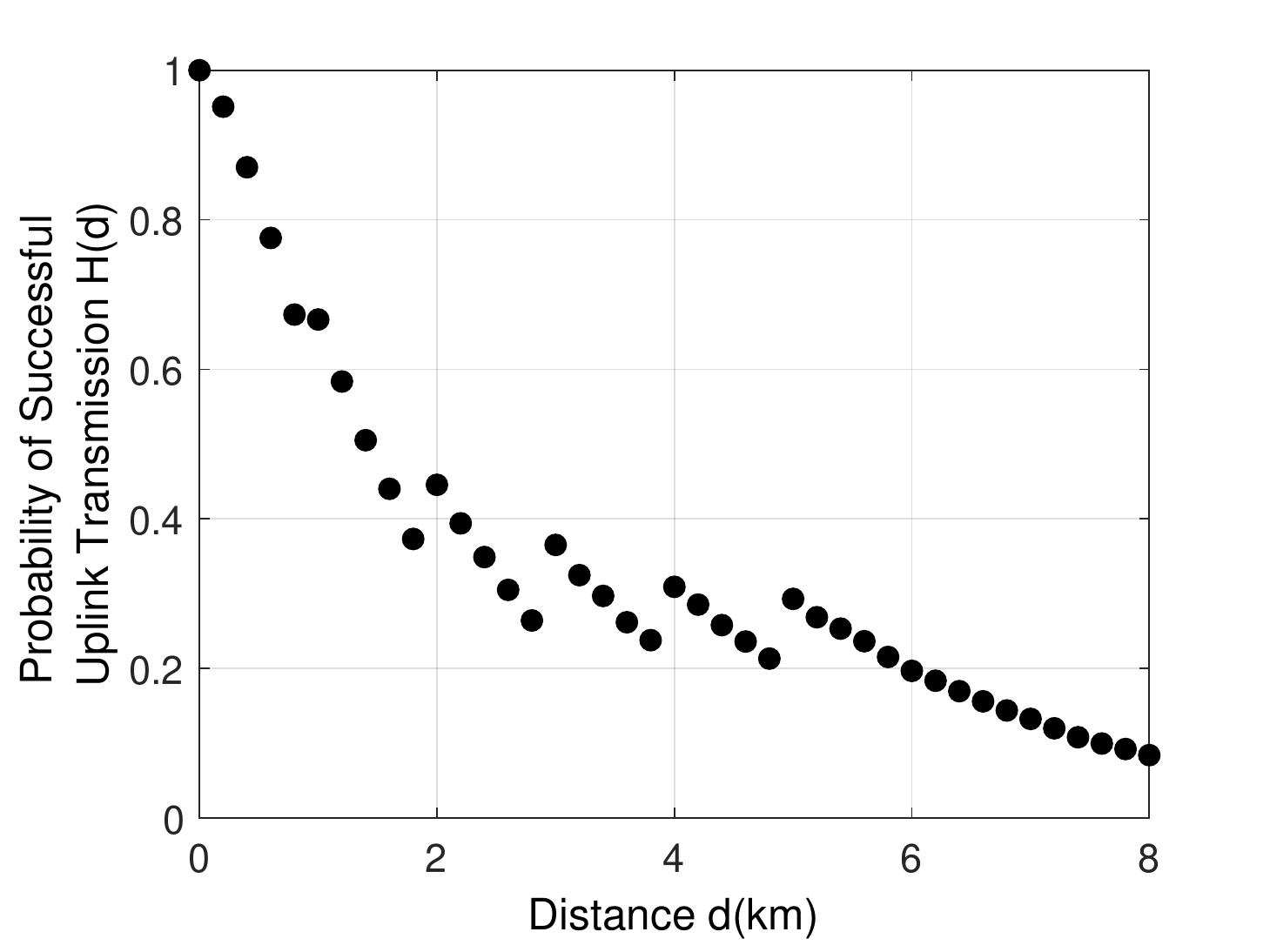}
  \caption{}
\end{subfigure}
\begin{subfigure}{0.5\textwidth}
  \includegraphics[width=1\linewidth]{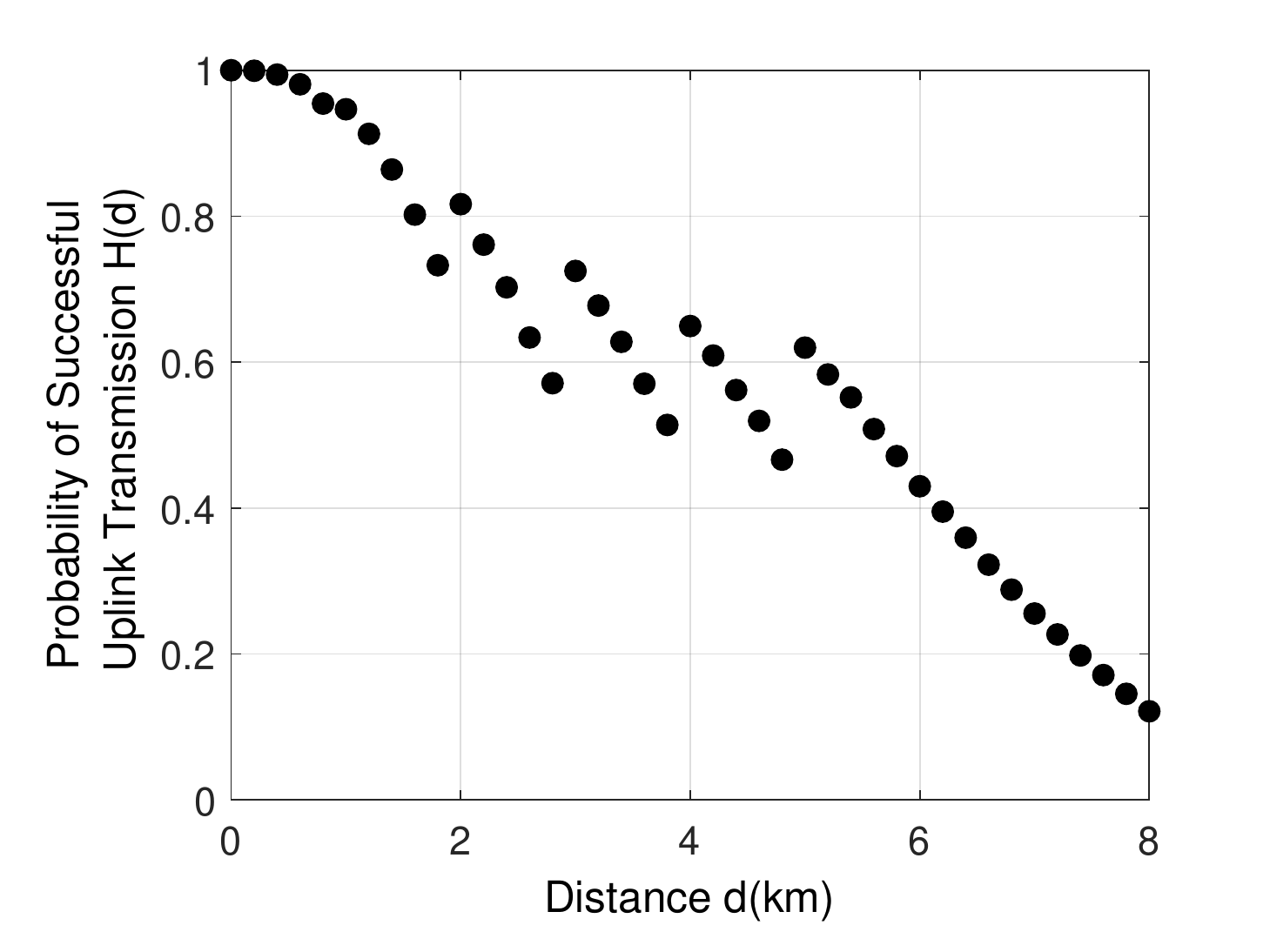}
  \caption{}
\end{subfigure}
\caption{Comparisons of numerical simulations for the success probability $H(x)$ between (a) Rayleigh channel and (b) Rician channel.}
\end{figure}

Observing from Fig.3b, we can clearly see that the Rician channel maintains a success rate of more than 50\% in the first 5 kilometers. On the other hand, in Fig.3a, Rayleigh's stats show that it can only keep that same number in the first 1.5 kilometers. The success rate decreases over distance for both channels. In both cases, EDs having $SF$ of 12 have a very low chance of successfully transmitting due to the vast number of co-SF interfering EDs. This is clearly understandable, since the further EDs are away from their nearest GW, the higher $SF$  they have, which leads to the fact that EDs are more likely to counter interference from other co-SF EDs.

\section{Conclusion}
\label{Sect:Conclusion}

In this paper, we investigated the performance and success probability of multi-cell LoRa networks. In such a system, interference caused by simultaneous transmissions using the same SF as well as different SFs is a problem that needs to be analyzed. To this end, we use the tool of stochastic geometry to analyze the important metric namely success probability in both Rayleigh and Rician channel models, while considering the impact of interference among transmissions over the same $SF$ (co-SF) as well as different $SFs$ (inter-SF). Moreover, we derive the important quality criterias such as throughput, SNR and SIR requirements in the multi-cell LoRa network. In addition, we summarized the network performance under Rayleigh and Rician channel models. 

\bibliographystyle{IEEEtran}
\bibliography{reference}
\begin{IEEEbiography}[{\includegraphics*[width=1in]{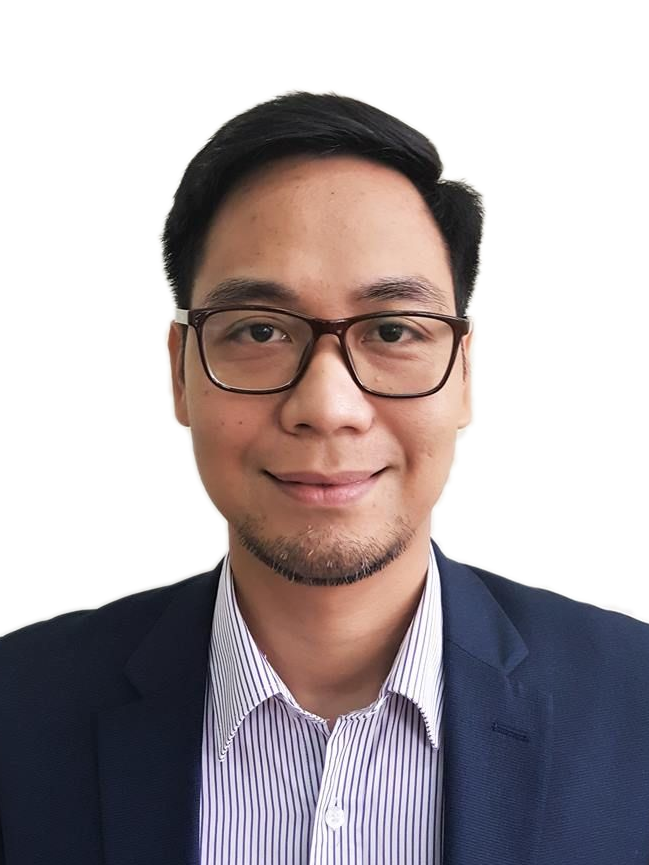}}]
{Nguyen Tien Hoa} graduated with a Dipl.-Ing. in Electronics and Communication Engineering from Hanover University. He has worked in the R\&D department of image processing and in the development of SDR-based drivers in Bosch, Germany. He devoted three years of experimentation with MIMOon's R\&D team to develop embedded signal processing and radio modules for LTE-A/4G. He worked for six months as a senior expert at Viettel IC Design Center (VIC) for development of advanced solutions for aggregating and splitting/steering traffic at the PDCP layer and above to provide robust and QoS/QoE guaranteeing integration between heterogeneous link types in 5G systems. Currently, he is a lecturer at the School of Electronics and Telecommunications, Hanoi University of Science and Technology. His research interests are resource allocation in B5G, and vehicular communication systems.
\end{IEEEbiography}
\begin{IEEEbiography}[{\includegraphics*[width=1in]{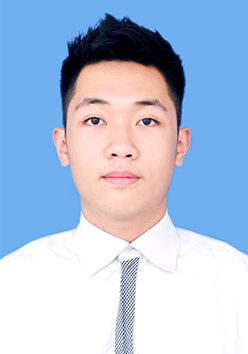}}]
{Van Dai Do} is a senior year student at Hanoi University of Science and Technology. He is following the program for the gifted in Electronics and Telecommunications. He spent two years working in the Technology Laboratory on Signal Processing. He also joined the talented program for Artificial Intelligence and Data Science hosted by Viettel Military Industry and Telecoms Group. His research interests are applying Machine Learning models to solve big data problems and signal processing in B5G/6G.
\end{IEEEbiography}
\vfill


\end{document}